%% file: main.tex
\documentclass[runningheads]{llncs}
\usepackage[T1]{fontenc}
\usepackage{graphicx}
\usepackage{algorithm}
\usepackage[noend]{algpseudocode}
\usepackage{xcolor}
\usepackage{xspace}
\usepackage[caption=false]{subfig}
\usepackage{enumitem}
\usepackage{pifont}
\usepackage{booktabs}
\usepackage[hidelinks]{hyperref}
\usepackage[capitalise]{cleveref}
\usepackage{siunitx}
\DeclareSIUnit{\us}{\SIUnitSymbolMicro s}

\newcommand{\tool}{Buddy\xspace}
\newcommand{\None}{\textsc{None}\xspace}
\newcommand{\Local}{\textsc{x86}\xspace}
\newcommand{\DPU}{\textsc{Dpu}\xspace}

\begin{document}
\title{Communication Offloading on SmartNIC DPUs: A Quantitative Approach}

\author{%
Jacob Wahlgren\inst{1} \and
Andong Hu\inst{1} \and
Roger Pearce\inst{2} \and
Maya Gokhale\inst{2} \and
Ivy Peng\inst{1}
}
\authorrunning{J. Wahlgren et al.}
\institute{%
KTH Royal Institute of Technology, Stockholm, Sweden \email{\{jacobwah,andonghu,ivybopeng\}@kth.se} \and
Lawrence Livermore National Laboratory, Livermore, USA \email{\{pearce7,gokhale2\}@llnl.gov}
}

\maketitle              %
\begin{abstract}
SmartNIC Data Processing Units (DPUs) offer a promising solution for saving high-end CPU resources by offloading tasks to programmable cores near the network interface. In this work, 
we explore the feasibility of SmartNIC DPUs in supporting an asynchronous communication model called ``fire-and-forget'', particularly its core message routing service. We design a communication offloading engine called Buddy that decouples communication tasks from the application process. Buddy runs flexibly on SmartNIC DPUs such as the Nvidia BlueField-3 DPU and generic x86 CPUs. Our evaluation results in five applications identify the memory-to-communication ratio as a key predictor of the offloading performance. Host-dominated workloads, such as Quicksilver and Sparse Matrix Transpose, achieved up to 1.55x speedup with communication offloaded to the DPU. We further identify a 625x increase in DRAM traffic due to the absence of Direct Cache Access support on the DPU, highlighting a critical need in future SmartNIC designs.

\end{abstract}

\input{intro2.tex}
\input{background.tex}

\input{design.tex}
\input{impl}
\input{setup.tex}
\input{results.tex}
\input{related.tex}

\input{conclusion.tex}

\subsubsection{Acknowledgment and Disclosures.}
This work was performed under the auspices of the U.S. Department of Energy by Lawrence Livermore National Laboratory under Contract 
DE-AC52-07NA27344 under LDRD Project 24-ERD-047. LLNL-CONF-2019409.
The authors have no competing interests to declare.

\bibliographystyle{splncs04}
\bibliography{main}
\end{document}

%% file: intro2.tex
\section{Introduction}
Data center networking is undergoing a transformation in the increasing adoption of SmartNICs to reduce resource cost and improve energy efficiency~\cite{tibbetts2025survey}. Unlike traditional passive Network Interface Cards (NICs), these emerging devices integrate fully-featured programmable CPUs, such as ARM and RISC-V cores, directly into the network interface, known as a Data Processing Unit (DPU). On-path SmartNICs feature embedded processors integrated directly in the network data path. In contrast, off-path SmartNICs offer decoupled, general-purpose processing, enabling offloading infrastructure tasks from the host CPU, freeing high-performance CPU cores to focus on application computation. This was previous demonstrated in MPI communication~\cite{bayatpour2021bluesmpi,suresh2023novel}, and workloads such as weather simulation~\cite{usman2025odos}, graph processing~\cite{wahlgren2024disaggregated} and molecular dynamics~\cite{karamati2022smarter}.

This work explores the feasibility of SmartNIC DPUs in supporting an asynchronous communication model called ``fire-and-forget''~\cite{ygm,maley2019conveyors,garg2006software,brock2019bcl}. This model aims to maximize parallelism in communication by relaxing message ordering and software aggregation and routing to bundle application-level messages into network-efficient packets. 
Traditionally, this routing service would co-run with application processes and share the compute and memory resources that could otherwise be used by the application.
In this work, we explore offloading the communication routing service to SmartNIC DPUs, identify key application metrics useful for configuring offloading, and draw design insights for future SmartNIC designs to support communication offloading efficiently. 

To quantitatively study offloading the routing service needed in the ``fire-and-forget'' model, we design \textit{Buddy},\footnote{\url{https://github.com/KTH-ScaLab/buddy}} a communication library that supports sending messages of arbitrary sizes asynchronously without involving the receiver. Buddy can either co-run on the host or be offloaded to the BlueField DPUs. When offloading message routing to the DPU, Buddy decouples communication management from application logic to reduce the resource usage on the host CPU. We use Buddy to quantitatively study key factors of communication offloading performance, such as task partitioning between DPU and CPU, aggregation of fine-grained messages to network-efficient packets, and SmartNIC architectural aspects.

In summary, the main contributions of this work are:
\begin{itemize}
    \item Design, implement, and optimize Buddy, a communication routing offloading library for ``fire-and-forget'' model suitable for SmartNIC DPUs and generic hosts.
    \item Evaluate performance in five applications: Single Source Shortest Path (SSSP), Triangle Counting, Histogram, Sparse Matrix Transpose, and Quicksilver. Quicksilver, Sparse Matrix Transpose, and Histogram achieve speedup from 1.13$\times$ to 1.55$\times$, with the memory-to-communication ratio as an effective indicator.
    \item Identify key optimization factors: transfer granularity, pipelining, and multi-threading, contributing to 9.5$\times$, 12.4$\times$, and 6.2$\times$ improvement. 
    \item Compare the BlueField-3 DPU against an x86 CPU for an offloading target and find the lack of DCA support as the main performance limitation for communication offloading.
\end{itemize}

%% file: background.tex
\section{Background}
\label{sec:bg}
\begin{figure}[bt]
    \centering
    \includegraphics[width=.75\linewidth]{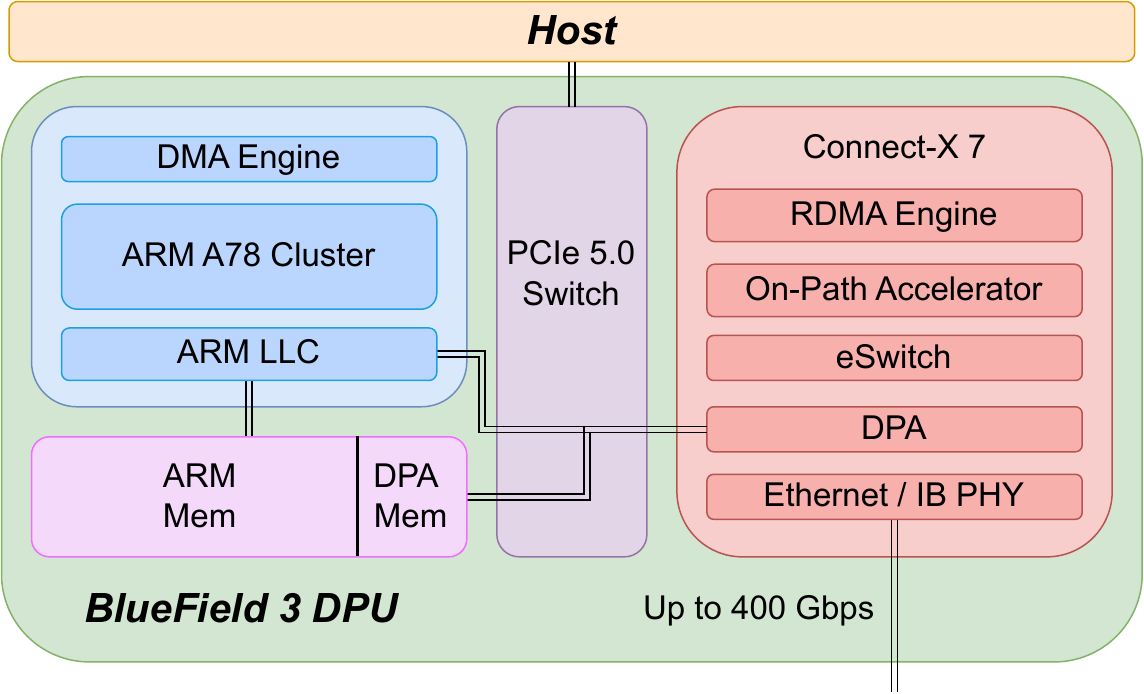}
    \caption{An overview of Nvidia BlueField-3 DPU architecture.}
    \label{fig:bf3arch}
\end{figure}

\textbf{SmartNICs \& DPUs.} SmartNICs and DPUs have emerged to address challenges in performance and efficiency. A SmartNIC extends the capability of a NIC by integrating programmable logic (e.g. ASIC/FPGA) on the data path, enabling host-side workloads to be offloaded to the NIC. Data Processing Units (DPUs) go one step further, featuring general-purpose CPUs and a standalone Linux software stack. %
Several vendors provide commercial DPU solutions with ARM cores, including AMD Pensando, Intel IPU, and Nvidia BlueField. %
In this work, we focus on the BlueField-3. \cref{fig:bf3arch} illustrates the overall architecture. BlueField-3 combines a Connect-X~7 NIC with an ARM~A78 CPU that consists of 8 to 16 cores. Compared to its predecessor BlueField-2, its memory system is upgraded to dual-channel 32~GiB LPDDR5 memory. The programming environment for BlueField is provided by Nvidia's Data-Center-on-a-Chip Architecture (DOCA) framework, which offers a set of libraries, drivers, and APIs that abstract low-level hardware resources such as DMA and RDMA engines.

\textbf{High-Performance Networking.} High-performance I/O devices generally use Direct Memory Access (DMA) to read and write directly to the main memory without involving CPU. In networking, this concept is extended to Remote Direct Memory Access (RDMA). With RDMA, nodes in a cluster can exchange data without CPU involvement. With one-sided RDMA, the initiator can READ or WRITE to specific memory addresses, while two-sided RDMA enables the specific addresses to be determined by the recipient with the SEND operation. As each operation incurs an overhead both in the NIC and the network switches, larger messages generally improve the throughput. Direct Cache Access (DCA), also known as cache injection, is a technique to improve I/O performance beyond DMA~\cite{leon2007reducing}. DCA enables network devices to directly inject data into the CPU's caches, bypassing the main memory~\cite{wang2022understanding,farshin2020reexamining}. %

%% file: design.tex
\section{\tool{} Design}
\subsubsection{Overview.}
\begin{figure}[bt]
    \centering
    \includegraphics[width=.75\linewidth]{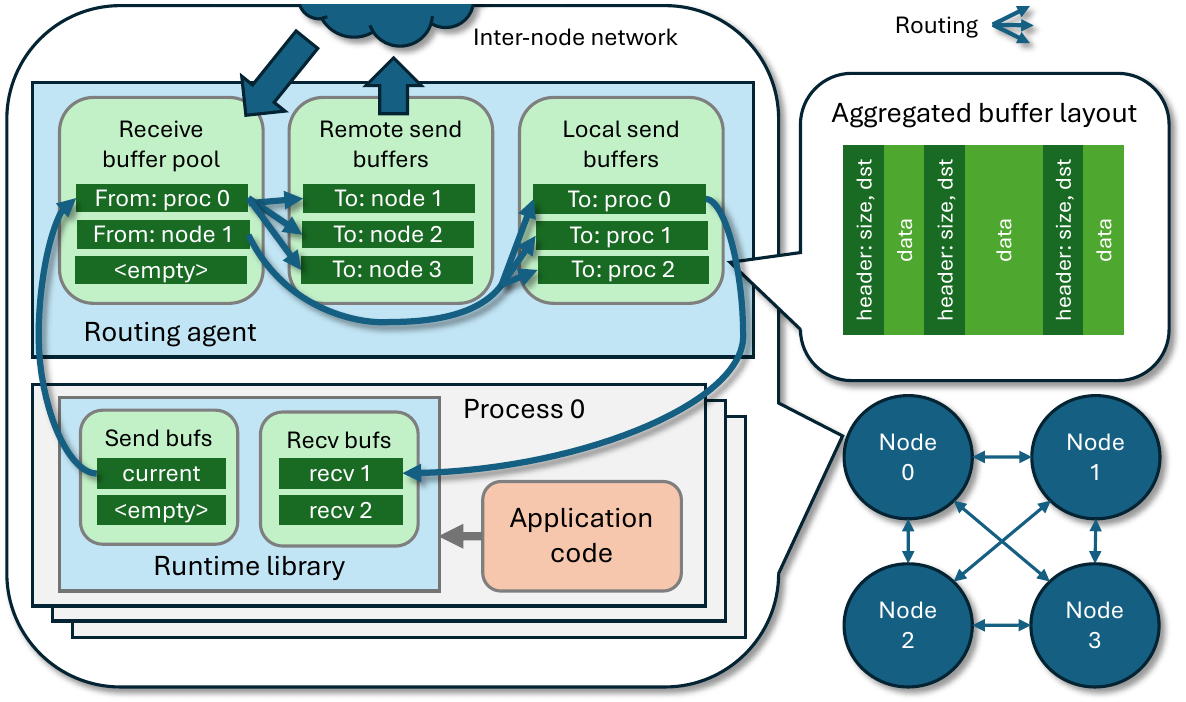}
    \caption{An overview of the \tool{} design, where the routing agent interfaces between local processes and remote nodes to aggregate individual messages.}
    \label{fig:design}
\end{figure}

Frequent irregular communication is a challenge in HPC systems since it leads to many small inefficient messages across the network. Message aggregation combined with multi-hop routing, a core component for enabling the ``fire-and-forget'' model, provides higher throughput by a trade-off with latency~\cite{ygm}. The design goal of Buddy is to decouple  message routing from the application, and reduce the burden on the host processor by offloading it to the DPU. Message routing services are memory-bound, and thus  may pollute cache and stall CPU cores for the core application work. Offloading routing services to DPUs frees the host processor to focus on the core application work instead of communication. Buddy targets distributed applications that exhibit the following characteristics:
\begin{itemize}
    \item Small and arbitrary messages. Processes communicate with messages significantly smaller than the link MTU.
    \item Fire-and-forget. Messages can be delivered asynchronously without requiring explicit confirmation from the receiver to the sender.
    \item Throughput-oriented. Application performance is limited by throughput rather than the latency of individual messages.
    \item Relaxed message ordering. Messages may be delivered out of order, enabling optimized routing and avoiding head-of-line blocking.
\end{itemize}

Our design is illustrated by \cref{fig:design} and consists of two major components: the \textit{runtime library} and the \textit{routing agent}. The runtime library offers an API for applications to interact with Buddy to send and receive messages, while the routing agent is a separate service handling the routing of incoming and outgoing messages between local application processes and remote cluster nodes. The routing agent can either co-run on the same CPU as the application, or be offloaded to another processor, e.g., a DPU.
\cref{fig:design} shows the different communication buffers and operations involved in sending and receiving messages. On the routing agent, three groups of communication buffers are used: receive buffers, remote send buffers, and local send buffers. These components make extensive use of zero-copy RDMA to avoid overheads.

Message aggregation happens at each hop in the design. First, the runtime library aggregates all outgoing messages, which could be of arbitrary sizes, from the application to ensure efficient transfer to the routing agent. Next, the local routing agent on the sending side aggregates all messages destined to the same remote node. Similarly, the routing agent on the receiving side aggregates incoming messages to the same local rank (i.e., Local send buffers in \cref{fig:design}). Finally, it delivers a set of aggregated messages from different sources to the destination process.

\textbf{Routing Agent.} The routing agent receives messages from local processes on the node and from remote nodes. As messages are received, the routing agent parses the message headers and routes them to their destination. The aggregated send buffers are flushed once they are full or a timeout is reached to avoid stale communication. To ensure that all messages are eventually delivered, the routing agent employs an idle timeout. If no messages are received for a certain time, the routing agent flushes all outstanding send buffers, even if they are not full.

\begin{algorithm}[bt]
    \caption{Routing kernel.}
    \label{alg:route}
    \begin{algorithmic}
    \Function{route}{array of aggregated messages}
        \For{message m}
            \State total\_size = sizeof(header) + m.payload\_size
            \State next = routing\_table[m.dst]
            \State send\_buf = get\_buf(next, total\_size)
            \If{send\_buf}
                \State memcpy(send\_buf.tail, m)
                \State send\_buf.tail += total\_size
            \Else
                \State \Return false
            \EndIf
        \EndFor
        \State \Return true
    \EndFunction
    \end{algorithmic}
\end{algorithm}

The routing kernel is the key performance-critical operation in the agent. A simplified version is shown in \cref{alg:route}. The kernel is applied to all incoming message bundles, both from local processes and from remote nodes. The time complexity of the routing algorithm is $O(n+m)$ for $n$ messages and $m$ total bytes, which is independent of the number of endpoints. We precompute the next-hop routing table at initialization. During routing, if no send buffer is available, the kernel exits prematurely, and the remaining messages are put on a blocklist. Once an outstanding send is completed, the kernel is applied to the entries of the blocklist.
Multiple threads are used in message routing to leverage the full hardware performance. Locks or atomics would lead to significant contention in each loop iteration. Instead, separate send buffers are allocated per thread.

\textbf{Runtime Library.} The runtime library offers support for applications to send and receive arbitrary sized messages with Buddy. It provides the necessary interfaces to bootstrap Buddy communications and manage aggregated send and receive buffers. All runtime library functions are non-blocking, enabling applications to maximize the time for useful computation. Furthermore, the runtime library leverages multiple send and receive buffers to enable pipelining of application compute with communication in a similar fashion to the routing agent. The low-level API consists of functions for sending or receiving aggregated message buffers and a call to poll for completions. %

\textbf{Communication Protocol.} Buddy communication travels two different data paths: intra-node between the runtime library and the routing agent, and inter-node between routing agents on different nodes. For inter-node communication, based on our characterization results on the testbed, RDMA is used for low-latency and high-bandwidth data movement for simplicity, portability, and flexibility. This enables the routing agent to be able to run both on a DPU or on any generic host cores. The RDMA protocol is based on SEND/RECV operations for its high performance since it avoids the need for distributed bookkeeping. Note that we also considered the DPU's DMA API, but benchmarking revealed no performance advantage over standard RDMA on the testbed. Message aggregation is implemented by concatenating multiple small messages in each RDMA request. %
Each message is encoded as an 8-byte header containing the payload size and destination rank, followed by the payload data. This format enables flexibility in the application as arbitrary-sized messages can be mixed freely.

%% file: impl.tex
\subsubsection{Implementation.}

We implement the Buddy routing agent and runtime library in C++. Communication is bootstrapped using MPI, where we use MPI ranks as addresses of application processes. RDMA communication is implemented with the \textit{ibverbs} library. All communication buffers are pre-allocated at initialization time to avoid the overhead of memory registration during runtime. In the routing agent, multi-threading is implemented by using OpenMP threads and a shared receive queue.

%% file: setup.tex
\section{Experimental Setup}
\label{sec:setup}

Our first testbed has two nodes equipped with BlueField-3 DPUs. The hosts have dual-socket Intel Xeon Gold 6326 CPUs, 16 cores each, and 512~GiB DDR4 memory. The DPUs have 8~ARM cores and 16~GiB LPDDR5 memory. The NIC is capable of 100~Gbps InfiniBand or RoCE link. The connection between hosts and DPUs is PCIe 4 x16, which has a maximum unidirectional bandwidth of 31.5 GB/s.
Both of the hosts and DPUs are running Linux 5.15 with DOCA version 2.7.0085. On this testbed, we used up to 8~threads for routing in all scenarios, which is the number of DPU cores. For scalability demonstration, we employ another testbed with 8 BlueField-3 nodes. The hosts have an AMD EPYC 9454 CPU with 48~cores and 770~GiB DDR5 memory. The DPUs have 16~ARM cores and 32~GiB LPDDR5 memory. This system also runs Linux 5.15 but has DOCA version 3.2.1. %

\begin{table}[bt]
    \centering
    \caption{Tested benchmarks and applications.}
    \begin{tabular}{lll}
        \toprule
        Application & Domain  & Comm. pattern\\
        \midrule
        Histogram & Data Analysis & Data-dependent \\
        Sparse Transpose & Linear Algebra & All-to-all \\
        Triangle Counting & Graph Analysis & Topology-dependent\\
        SSSP & Graph Analysis & Topology-dependent \\
        QuickSilver & Physics & 3D neighborhood \\
        \bottomrule
    \end{tabular}
    \label{tab:apps}
\end{table}

We use a set of five applications to evaluate communication offloading. They come from a variety of domains and feature different communication patterns as shown in \cref{tab:apps}.
Each of the applications features irregular communication patterns as they send fine-grained asynchronous messages in an unstructured fashion. The Histogram, Sparse Transpose, Triangle Counting, and SSSP (Single Source Shortest Path) benchmarks are based on~\cite{maley2019conveyors} for which we developed a compatability layer. For these applications, we use randomized inputs and set the problem size to 100x the host's last-level cache. Quicksilver is a proxy application that solves a dynamic Monte Carlo particle transport problem. We ported it to the low-level API with around 200 inserted and 400 deleted lines. For the experiments, we use the CTS2 benchmark problem with a reduced number of steps. %

\begin{figure}[bt]
    \centering
    \subfloat[\None]{%
        \includegraphics[width=0.28\linewidth]{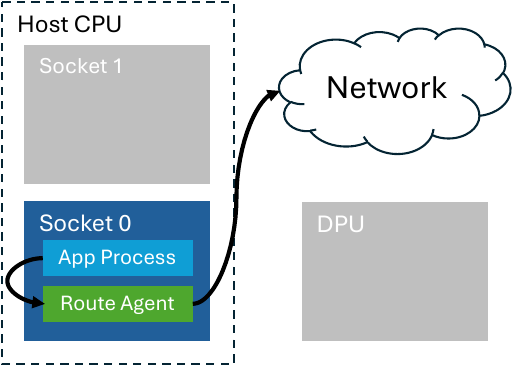}%
    }\hfill
    \subfloat[\Local]{%
        \includegraphics[width=0.28\linewidth]{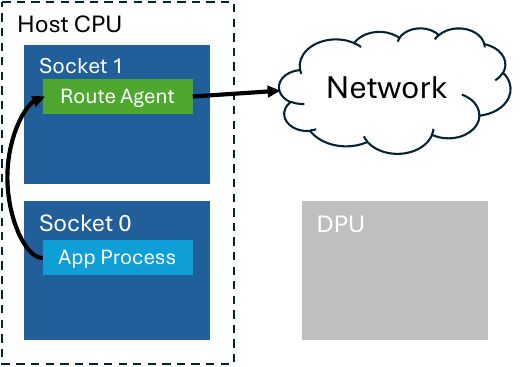}%
    }\hfill
    \subfloat[\DPU]{%
        \includegraphics[width=0.28\linewidth]{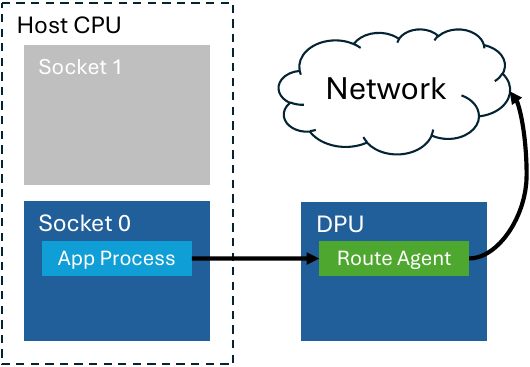}%
    }
    \caption{Three offloading scenarios with arrows representing RDMA data transfer.}
    \label{fig:offload-modes}
\end{figure}

We compare three offloading scenarios in this work, as illustrated by \cref{fig:offload-modes}. Since Buddy enables communication offloading using a generic infrastructure that is portable to different environments, this enables us to compare the routing offloading performance with the routing agent deployed on different hardware. The first non-offloading scenario does not decouple the routing component from application. The second and third scenarios decouple the routing agent either to a separate socket or the SmartNIC DPU.
\begin{itemize}
    \item \None: The application process and routing agent both run on the same host cores (with simultaneous multi-threading) as a performance baseline.
    \item \Local: The routing agent is offloaded to a separate socket with a powerful host CPU, as an upper bound on communication offloading performance.
    \item \DPU: The routing agent is offloaded to low-power cores on the SmartNIC.
\end{itemize}

%% file: results.tex
\section{Application Performance}

\newcommand{\cmark}{\ding{51}\xspace}
\newcommand{\xmark}{\ding{55}\xspace}

\subsubsection{Offloading Performance.}
\begin{figure}[bt]
    \begin{minipage}[t]{0.31\linewidth}
        \centering
        \includegraphics[width=\linewidth]{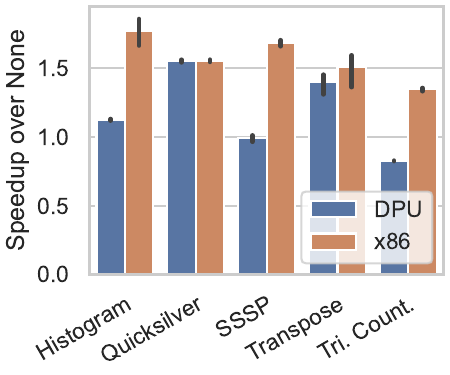}
        \caption{Speedup compared to no offloading.}
        \label{fig:overall}
    \end{minipage}\hfill
    \begin{minipage}[t]{0.31\linewidth}
        \centering
        \includegraphics[width=\linewidth]{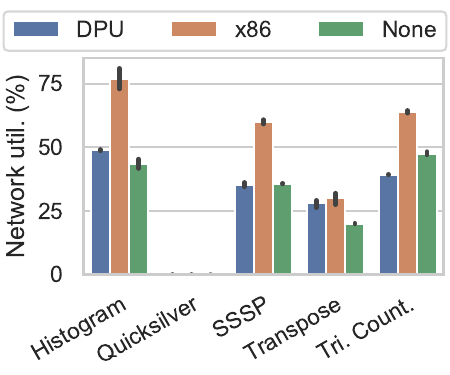}
        \caption{Network utilization in different scenarios.}
        \label{fig:netperf}
    \end{minipage}\hfill
    \begin{minipage}[t]{0.31\linewidth}
        \centering
        \includegraphics[width=\linewidth]{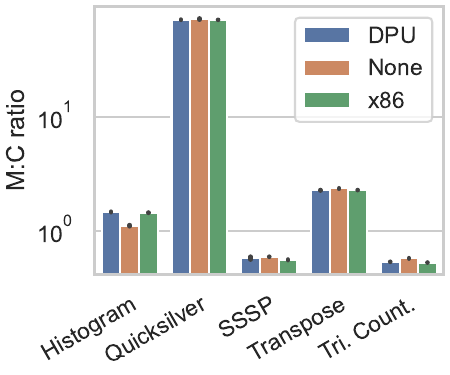}
        \caption{M:C ratio in different scenarios (log scale).}
        \label{fig:comm_ratio}
    \end{minipage}%
\end{figure}

A comparison of the application-level performance in the different offloading scenarios is shown in \cref{fig:overall}. The plot shows the mean performance of five runs.
In all applications, \Local offloading improved performance compared to the non-offloading baseline, from a mean of 1.35$\times$ speedup in Triangle Counting to 1.77$\times$ in Histogram. \DPU offloading improved performance in three applications, with 1.13$\times$ speedup in Histogram, 1.40$\times$ in Sparse Transpose, and 1.55$\times$ in Quicksilver. The performance was not improved in SSSP and Triangle Counting, with 0.99$\times$ and 0.83$\times$ speedup, respectively.

To explain the performance difference between the applications, we employed additional profiling metrics. First, we measured the network utilization, shown in \cref{fig:netperf}. We calculate it as the inter-node communication throughput of Buddy over the theoretical link speed.
In the \Local scenario, Histogram reached a high network utilization of 77\%. SSSP and Triangle Counting also achieved high utilizations with 60\% and 64\%, respectively. These applications also had the largest difference between the \Local and \DPU performance. Evidently, they can benefit from communication offloading, but \DPU offloading is limited by bottlenecks in the DPU hardware, i.e., weaker cores and memory.
Two applications had a lower network utilization, with 30\% for Sparse Transpose and only 0.1\% for Quicksilver.
They have more local work relative to the amount of communication, and therefore benefit from offloading as it frees up resources for local computation work. Offloading to x86 cores is more beneficial because it suffers less from the weaker DPU system compared to \Local.

While tasks such as Sparse Transpose contain no floating-point computations, traversing the sparse structure incurs many cache misses and constitutes the major local work in the benchmark.
We measured the ratio of memory accesses to communication (M:C ratio) in the applications to characterize the balance of local work to communication. The total amount of memory accessed is measured as the number of LLC cache misses times the cache line size (with HPCToolkit), while the amount of communication is measured as the number of bytes routed by the routing agent. For instance, a ratio of 2 means that the application loads 2 bytes from memory for each 1 byte it communicates with Buddy.
The M:C ratios are shown in \cref{fig:comm_ratio}. They are largely consistent across the offloading scenarios, although it deviates slightly for \None in Histogram.
Quicksilver has by far the highest ratio around 72, explaining the low network utilization we observed,
followed by Sparse Transpose with 2.3 and Histogram with 1.5. SSSP and Triangle Counting have more communication than memory traffic with ratios of 0.56 and 0.53, respectively.

\begin{table}[bt]
    \centering
    \caption{Summary of application offloading results.}
    \label{tab:summary}
    \begin{tabular}{lccccl}
        \toprule
        Application & Speedup \Local & Speedup \DPU & Net. util. & M:C ratio & Category \\
        \midrule
        Histogram & \cmark (1.77) & \cmark (1.13) & 77\% & 1.5 & Balanced \\
        Quicksilver & \cmark (1.55) & \cmark (1.55) & 0.1\% & 72 & Host-dominated \\
        SSSP & \cmark (1.68) & \xmark (0.99) & 60\% & 0.56 & Comm-dominated \\
        Transpose & \cmark (1.51) & \cmark (1.40) & 30\% & 2.3 & Host-dominated \\
        Tri. Count & \cmark (1.35) & \xmark (0.83) & 64\% & 0.53 & Comm-dominated \\
        \bottomrule
    \end{tabular}
\end{table}

We summarize the results in \cref{tab:summary}. Based on the M:C ratio, we categorize the applications into three groups. Quicksilver and Sparse Transpose have a high degree of local work (M:C ratio $>2$) and are therefore \textit{host-dominated}. SSSP and Triangle Counting have a high degree of communication (M:C ratio $<1$) and are therefore \textit{communication-dominated}. Histogram lies in between, we consider it \textit{balanced}. The host-dominated applications see a significant speedup (1.40-1.55) from \DPU offloading, while the communication-dominated applications see no benefit. The balanced Histogram has a modest speedup of 1.13.

\begin{figure}[bt]
    \centering
    \includegraphics[width=\linewidth]{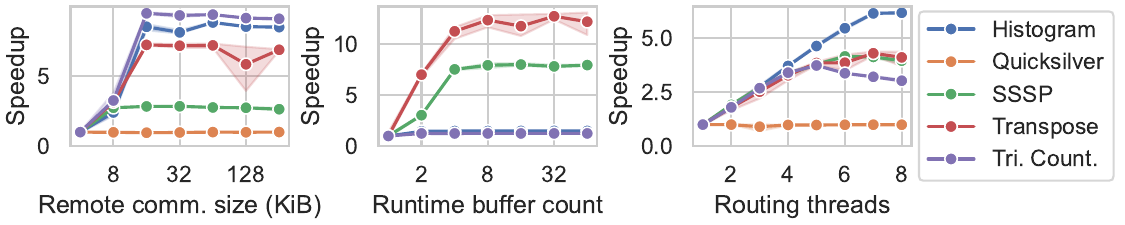}
    \caption{Impact of tuning and optimizations on application performance.}
    \label{fig:sweep}
\end{figure}

\subsubsection{Optimizations.}
We studied the impact of tuning and optimizations on communication offloading performance by measuring application performance with varied Buddy configurations in \DPU mode. Remote communication size is the size of buffers for routing-agent to routing-agent traffic over the network, with 4~KiB (i.e. the link MTU) as the baseline. Runtime buffer count is the number of send and receive buffers in the runtime library. Routing threads is the number of OpenMP threads in the routing agent. The speedup is shown in \cref{fig:sweep}.

In most cases, increasing the communication size to 8~KiB results in a 2-3x speedup, reaching a plateau around 16-32~KiB, with up to 9.5x in Triangle Counting, due to increased communication efficiency.
The performance impact on Quicksilver is negligible, as it is strongly host-dominated.
Increasing the runtime buffers enables communication pipelining and better overlap. The performance of SSSP and Transpose was significantly improved, reaching a plateau around 8~buffers with 8.0x and 12.4x, respectively, while other applications were not significantly affected.%
Utilizing multiple threads in the routing agent improves performance for most applications, with up to 6.2x speedup at 7 or 8 threads in Histogram. Triangle Counting peaks at 5~threads with 3.7x speedup, while SSSP and Tranpose peak at 7~threads with 4.2x and 4.3x, respectively.
These results show that communication granularity, pipelining, and multi-threaded DPU services are key optimizations for communication offloading.

\begin{figure}[bt]
    \centering
    \includegraphics[width=.85\linewidth]{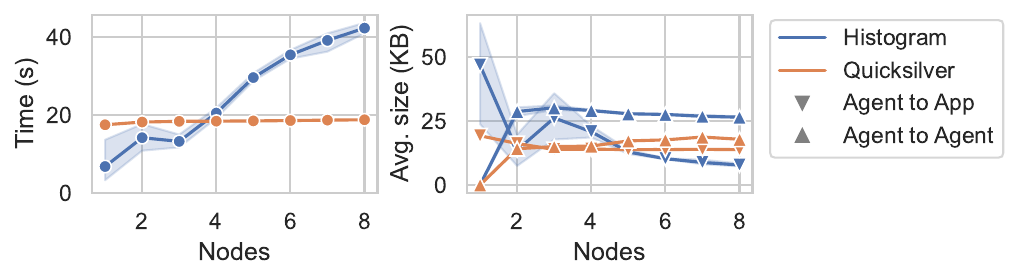}
    \caption{Performance of scaling Histogram and Quicksilver with \DPU offloading up to 8 nodes and the average transfer sizes.}
    \label{fig:scaling}
\end{figure}

\subsubsection{Scalability.} We evaluated the scalability of Buddy communication offloading on the 8-node cluster by performing a weak scaling test of Histogram and Quicksilver with \DPU offloading. In Histogram, we scale both the size of the distributed table and the total number of updates, and in Quicksilver, we scale both the number of mesh elements and the number of particles. The routing agent uses 16 threads. The execution time and average transfer size per run across 5 runs are shown in \cref{fig:scaling}.

Quicksilver achieves close to ideal weak scaling with almost constant execution time regardless of the number of nodes. This is in line with our finding that Quicksilver is host-dominated.
In Histogram, the execution time is fastest with 1 node, as there is no cost for inter-node data transfer. With 2 and 3 nodes, the time is similar. From 3 to 5 nodes, the time scales linearly, after which the growth rate decreases slightly.

The average transfer size highlights a factor limiting the scalability of Histogram. While the average agent-to-agent (i.e. cross-node) transfer size remains in the range 26-30~KB (for nodes $>1$), the average agent-to-application (i.e. intra-node) transfer size decreases from 26~KB at 3 nodes to 8~KB at 8~nodes.
On the other hand, the transfer size in Quicksilver remains in the range 14-19~KB for all distributed cases.
These findings indicate that Buddy's scalability on balanced or communication-dominated workloads could be improved by further tuning buffer sizes and idle timeouts for agent-to-application transfers on multiple nodes.

\section{DPU Bottleneck Analysis}

\subsubsection{BlueField-3.}
To understand how DPU design can be enhanced to support communication offloading, we analyze the system bottlenecks.
In the previous section, we observed that \DPU offloading is weaker than \Local offloading, which could be explained by weaker cores, memory subsystem, or higher data transfer costs in the DPU scenario. However, we cannot tell from these experiments which system component is the bottleneck.

We investigate two potential bottlenecks by injecting interference and measuring the relative performance of Buddy. As a baseline case, we use \DPU offloading with 7~threads. To inject DPU memory interference, we concurrently execute a memory-intensive workload on the 8th core. To inject interference into the link between the host and DPU, we concurrently execute a communication-intensive workload on the 8th core and a free core on the host. The memory-intensive workload is a vector addition with a working set of 32~MB (more than 2x the last-level cache capacity). The communication-intensive workload is the \verb|ib_send_bw| RDMA benchmark in bidirectional mode with 64~KB message size.
For these experiments, we use the Histogram application, which has a large disparity between \Local and \DPU performance, and repeat experiments five times.

\begin{figure}[bt]
    \centering
    \begin{minipage}{0.31\linewidth}
        \includegraphics[width=\linewidth]{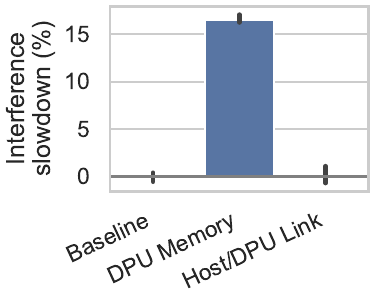}
        \caption{Performance of \DPU offloading with different kinds of interference.}
        \label{fig:bottleneck}
    \end{minipage}\hfill
    \begin{minipage}{0.31\linewidth}
        \includegraphics[width=\linewidth]{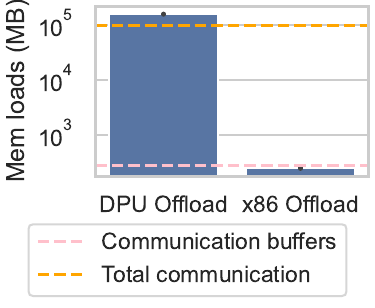}
        \caption{Amount of data loaded from DRAM in the routing agent (log scale).}%
        \label{fig:mem_traffic}
    \end{minipage}\hfill
    \begin{minipage}{0.31\linewidth}
        \includegraphics[width=\linewidth]{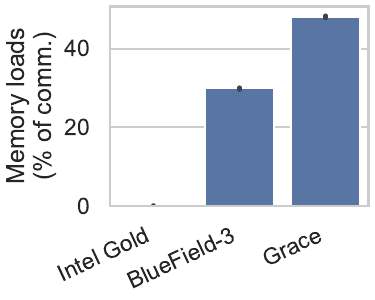}
        \caption{LLC misses fraction of total network traffic in micro-benchmark.}%
        \label{fig:dca}
        \end{minipage}
\end{figure}

\cref{fig:bottleneck} presents the slowdown of the application with different sources of interference. Communication interference does not significantly impact the performance of \DPU offloading. However, DPU memory interference decreases the performance by 17\%, indicating that the DPU's main memory is the performance bottleneck for \DPU offloading. Furthermore, we profile LLC misses in the routing agent using \texttt{perf} to compare the frequency of DRAM accesses in \DPU and \Local offloading, as shown in \cref{fig:mem_traffic}. The total L3 cache misses correspond to 249~MiB and 153~GiB of data loaded from memory in \Local and \DPU, respectively. The value for \Local roughly corresponds to reading all of the communication buffers a single time (272~MiB), while the value for \DPU is 1.6x the total communicated data volume (95~GiB). The ratio of cache misses corresponds to a 625x increase in main memory demand loads with \DPU offloading compared to \Local offloading. The low memory traffic in \Local can be explained by the fact that the CPU supports Intel DDIO, and thus utilizes DCA to place arriving data directly into the cache without going through main memory. While ARM has a DCA feature called Cache Stashing, the larger memory traffic in \DPU indicates that it is not supported or not enabled on the BlueField-3 ARM cores.

In conclusion, the DPU memory is the bottleneck in Buddy applications with \DPU offloading. Memory traffic is significantly higher in \DPU than \Local because the DPU does not leverage DCA.

\subsubsection{Future SmartNICs.}
\label{sec:bf4}

Addressing the identified bottleneck in future SmartNICs would enable DPU-based communication processing, such as the software routing, closer to the underlying network speed, benefiting high-throughput DPU applications.
At the time of writing, Nvidia recently announced its next-generation SmartNIC BlueField-4. %
We note that the BlueField-4 will leverage the Nvidia Grace CPU for the ARM cores of the DPU and a more powerful memory system, which could address the identified limitations of weak CPU and memory in the previous generations of BlueField DPUs for enabling efficient routing offloading needed for the ``fire-and-forget'' model.

To inform expectations for the future SmartNIC DPU on the DCA capabilities, we designed an experiment to test on the Grace CPU. We utilized the \verb|ib_send_bw| benchmark from perftest, with a small modification to read each byte of received data on the main CPU thread instead of just discarding it. To test the DCA capabilities of a system, we execute the benchmark and measure the total number of LLC misses using \verb|perf stat|, and calculate the total cache misses as a fraction of the total amount of data received. Specifically, we set the message size to 64~KiB and the number of iterations to 10,000, giving a total of 625~MiB data, and scale the cache misses by the cache line size (64~B). We execute the benchmark in loopback mode.

We execute the benchmark on a system with an Nvidia Grace CPU and an Nvidia ConnectX-7 NIC.
Our Grace system features 72 ARM Neoverse~V2 cores and 480~GiB of LPDDR5X memory.
For comparison, we also execute the benchmark on Intel Gold and BlueField-3, corresponding to the \Local and \DPU offloading scenarios. The results from three runs are shown in \cref{fig:dca}. As expected, the Intel system has essentially no cache misses, while the BlueField-3 system has a significant amount with 30\%. Interestingly, the Grace system has an even larger amount of misses at 48\%. This indicates that the Grace system, like the BlueField-3, is not utilizing DCA. We hypothesize that CPU prefetching is the reason total cache misses on BlueField-3 and Grace remain below 100\%. These results show that DCA does not come ``for free'' by upgrading the CPU. To enable high-throughput communication offloading, future SmartNIC architects should consider DCA as a critical feature.

%% file: related.tex
\section{Related Works}
In HPC, SmartNICs have been used both for communication and as general-purpose coprocessors.
In \cite{bayatpour2021bluesmpi,suresh2023novel}, the MPI progress engine is offloaded to enable computation overlap while PEDAL~\cite{li2024accelerating} offloads in-flight compression and decompression in MPICH
Works such as~\cite{karamati2022smarter} offload specific application kernels, while
ODOS~\cite{usman2023dpu,usman2025odos} enables general-purpose offloading with OpenMP.
Outside the SmartNIC area, SDT~\cite{mamandipoor2025sdt} enhances CPU architecture with specialized Simultaneous Data-delivery Threads to co-run application threads and communication threads with performance isolation. %

Previous works have shown the benefits of software aggregation and routing in irregular communication without leveraging SmartNIC offloading.
Garg and Sabharwal~\cite{garg2006software} studied the  HPCC RandomAccess benchmark and demonstrated how message aggregation and software routing improved the performance by 2x on the Blue Gene/L supercomputer.
Subsequent works have proposed libraries and multiple routing and aggregation schemes, including %
BCL~\cite{brock2019bcl},
Conveyors~\cite{maley2019conveyors}, and YGM~\cite{ygm}.

One previous work claimed that DCA is already supported in BlueField-2 DPU~\cite{gu2024omniccl}. The authors draw this conclusion from the fact that the \verb|ib_write_bw| benchmark can achieve a bandwidth higher than the STREAM benchmark on the DPU. However, we believe that the conclusion is incorrect. Without modifying the benchmark, as we did in \cref{sec:bf4}, the CPU does not access the data at all, and the benchmark's throughput is limited only by the NIC's DMA capabilities rather than the CPU's memory bottleneck.

Other recent works on DCA only focus on Intel DDIO.
Farshin et al.~\cite{farshin2020reexamining} dissected the implementation of DDIO, proposed a set of optimization guidelines for performance isolation, and demonstrated DDIO fine-tuning. 
Alian et al.~\cite{alian2020data} presented a gem5 simulator model for DDIO and characterized its impact on memory utilization and performance on storage and network processing benchmarks.
Wang et al.~\cite{wang2022understanding} further studied the architectural details of DDIO to develop analytical models to diagnose performance issues, tune configurations, and inspire future hardware designs.

%% file: conclusion.tex
\section{Conclusion}

In this work, we explored the prospects of offloading communication routing with SmartNICs, a key component for enabling the ``fire-and-forget'' model. To this end, we developed a prototype offloading engine called Buddy and evaluated it on Nvidia BlueField-3. We found that DPU offloading efficiency depends on the amount of local work in the application. Specifically, the memory-to-communication ratio is a key indicator of DPU offloading performance, with host-dominated workloads gaining the most from offloading. Secondly, we identified the absence of Direct Cache Access (DCA) for network packets as an architectural bottleneck in the DPU, causing a 625x increase in memory traffic compared to an x86-based scenario with DCA, which highlights an important consideration for future SmartNIC designs.

%% file: main.bbl
\begin{thebibliography}{10}
\providecommand{\url}[1]{\texttt{#1}}
\providecommand{\urlprefix}{URL }
\providecommand{\doi}[1]{https://doi.org/#1}

\bibitem{alian2020data}
Alian, M., Yuan, Y., Zhang, J., Wang, R., Jung, M., Kim, N.S.: Data direct {I/O} characterization for future {I/O} system exploration. In: 2020 IEEE International Symposium on Performance Analysis of Systems and Software (ISPASS) (2020)

\bibitem{bayatpour2021bluesmpi}
Bayatpour, M., Sarkauskas, N., Subramoni, H., Maqbool~Hashmi, J., Panda, D.K.: {BluesMPI}: Efficient {MPI} non-blocking {Alltoall} offloading designs on modern {BlueField} smart {NICs}. In: International Conference on High Performance Computing (2021)

\bibitem{brock2019bcl}
Brock, B., Bulu{\c{c}}, A., Yelick, K.: {BCL}: A cross-platform distributed data structures library. In: Proceedings of the 48th International Conference on Parallel Processing (2019)

\bibitem{farshin2020reexamining}
Farshin, A., Roozbeh, A., Maguire~Jr, G.Q., Kosti{\'c}, D.: Reexamining direct cache access to optimize {I/O} intensive applications for multi-hundred-gigabit networks. In: 2020 USENIX Annual Technical Conference (USENIX ATC 20) (2020)

\bibitem{garg2006software}
Garg, R., Sabharwal, Y.: Software routing and aggregation of messages to optimize the performance of {HPCC} {Randomaccess} benchmark. In: Proceedings of the 2006 ACM/IEEE conference on Supercomputing (2006)

\bibitem{gu2024omniccl}
Gu, T., Fei, J., Canini, M.: {OmNICCL}: Zero-cost sparse {AllReduce} with direct cache access and {SmartNICs}. In: Proceedings of the 2024 SIGCOMM Workshop on Networks for AI Computing (2024)

\bibitem{karamati2022smarter}
Karamati, S., Hughes, C., Hemmert, K.S., Grant, R.E., Schonbein, W.W., Levy, S., Conte, T.M., Young, J., Vuduc, R.W.: {“Smarter”} {NICs} for faster molecular dynamics: a case study. In: 2022 IEEE International Parallel and Distributed Processing Symposium (IPDPS) (2022)

\bibitem{leon2007reducing}
Le{\'o}n, E.A., Ferreira, K.B., Maccabe, A.B.: Reducing the impact of the memory wall for {I/O} using cache injection. In: 15th Annual IEEE Symposium on High-Performance Interconnects (HOTI 2007) (2007)

\bibitem{li2024accelerating}
Li, Y., Kashyap, A., Chen, W., Guo, Y., Lu, X.: Accelerating lossy and lossless compression on emerging bluefield dpu architectures. In: 2024 IEEE International Parallel and Distributed Processing Symposium (IPDPS). pp. 373--385 (2024)

\bibitem{maley2019conveyors}
Maley, F.M., DeVinney, J.G.: Conveyors for streaming many-to-many communication. In: 2019 IEEE/ACM 9th Workshop on Irregular Applications: Architectures and Algorithms (IA3) (2019)

\bibitem{mamandipoor2025sdt}
Mamandipoor, A., Tran, H.D., Alian, M.: {SDT}: Cutting datacenter tax through simultaneous data-delivery threads. IEEE Computer Architecture Letters  (2025)

\bibitem{ygm}
Steil, T., Reza, T., Priest, B., Pearce, R.: Embracing irregular parallelism in {HPC} with {YGM}. In: Proceedings of the International Conference for High Performance Computing, Networking, Storage and Analysis (2023)

\bibitem{suresh2023novel}
Suresh, K.K., Michalowicz, B., Ramesh, B., Contini, N., Yao, J., Xu, S., Shafi, A., Subramoni, H., Panda, D.: A novel framework for efficient offloading of communication operations to {BlueField} {SmartNICs}. In: 2023 IEEE International Parallel and Distributed Processing Symposium (IPDPS) (2023)

\bibitem{tibbetts2025survey}
Tibbetts, N., Ibtisum, S., Puri, S.: A survey on heterogeneous computing using {SmartNICs} and emerging data processing units. Future Generation Computer Systems  (2025)

\bibitem{usman2025odos}
Usman, M., Benito, M., Iserte, S., Pe{\~n}a, A.J.: {ODOS-MPI}: {HPC}-friendly {SmartNIC} offloading of computation/communication kernels. In: Proceedings of the International Conference for High Performance Computing, Networking, Storage and Analysis (2025)

\bibitem{usman2023dpu}
Usman, M., Iserte, S., Ferrer, R., Pe{\~n}a, A.J.: {DPU} offloading programming with the {OpenMP API}. In: Proceedings of the SC'23 Workshops of The International Conference on High Performance Computing, Network, Storage, and Analysis (2023)

\bibitem{wahlgren2024disaggregated}
Wahlgren, J., Schieffer, G., Gokhale, M., Pearce, R., Peng, I.: Disaggregated memory with smartnic offloading: a case study on graph processing. In: 2024 IEEE 36th International Symposium on Computer Architecture and High Performance Computing (SBAC-PAD). pp. 159--169. IEEE (2024)

\bibitem{wang2022understanding}
Wang, M., Xu, M., Wu, J.: Understanding {I/O} direct cache access performance for end host networking. Proceedings of the ACM on Measurement and Analysis of Computing Systems  \textbf{6}(1) (2022)

\end{thebibliography}
